\documentclass{article}

\usepackage{ijcai21}

\usepackage{times}

\usepackage[utf8]{inputenc} 
\usepackage[T1]{fontenc}    
\usepackage{hyperref}       
\usepackage{url}            
\usepackage{booktabs}       
\usepackage{amsfonts}       
\usepackage{nicefrac}       
\usepackage{microtype}      
\usepackage{graphicx}
\usepackage{xcolor}
\usepackage{amsmath}

\newcommand{\citet}[1]{\citeauthor{#1}~\shortcite{#1}}
\newcommand{\citep}{\cite}

\title{A Synchronized Action Framework for Responsible Detection of Coordination on Social Media}

\author{
Thomas Magelinski\footnote{Contact Author}\and
Lynnette Hui Xian Ng\and
Kathleen M. Carley\\
\affiliations
Carnegie Mellon University, 5000 Forbes Avenue, Pittsburgh, PA, 15217 \\
\emails
\{tmagelin, huixiann, kathleen.carley\}@cs.cmu.edu
}

\begin{document}

\maketitle

\begin{abstract}
The study of coordinated manipulation of conversations on social media has become more prevalent as social media's role in amplifying misinformation, hate, and polarization has come under scrutiny. 
We discuss the implications of successful coordination detection algorithms based on shifts of power, and consider how responsible coordination detection may be carried out through synchronized action.
We then propose a Synchronized Action Framework for detection of automated coordination through construction and analysis of multi-view networks. 
We validate our framework by examining the Reopen America conversation on Twitter, discovering three coordinated campaigns.
We further investigate covert coordination surrounding the protests and find the task to be far more complex than examples seen in prior work, demonstrating the need for our multi-view approach.
A cluster of suspicious users is identified and the activity of three members is detailed.
These users amplify protest messages using the same hashtags at very similar times, though they all focus on different states.
Through this analysis, we emphasize both the potential usefulness of coordination detection algorithms in investigating amplification, and the need for careful and responsible deployment of such tools.

\end{abstract}

\section{Introduction}
Coordinated influence operations pose unique threats to social cybersecurity.
While people can be trained to recognize accounts that are likely bots, coordinated users may be inconspicuous until compared to other accounts.
The United States has previously seen political protests orchestrated by external actors unbeknownst to the protest's attendees, indicating that covert coordination could play a role in inciting protests \cite{weiss_2018}.
Inorganic coordination between communities threaten the social fabric of society with potential for offline violence.
Some examples are the coordinated efforts to propagate extremist ideas by ISIS, and coordinated efforts to share false stories about conflict during the Black Panther movie event \cite{benigni2017online,babcock2018beaten}. 

Previous works have explored a number of behaviors that may be coordinated: user profile features, tweet content, and activity patterns \cite{nizzoli2020coordinated,pacheco2020uncovering,vargas2020detection}.
In these studies, synchronized behaviors are only a special case of coordination.
A network-based approach is then taken, where users are connected in proportion to the number of times they exhibited the same behavior. 
These networks are typically constructed using a fixed-window, which, while easy to implement, results in missing connections.
In fact, under a uniform distribution of behaviors, a fixed-window approach will only find 50\% of the connections between users. 
Additionally, these networks are analyzed independently, providing only a singular perspective of the coordination activity.
The present work is concerned with network approaches, though non-network based approaches have been explored as well, using techniques like temporal point processes to model account activities \cite{sharma2020identifying}.

In-depth qualitative analysis plays a crucial role in understanding information operations.
However, synchronized behavior is extremely difficult to find with the human eye.
Automated detection of these activities can find potentially coordinated users, giving qualitative researchers additional signals for their analysis.

While the automated detection of coordinated accounts remains an important problem, some  solutions carry the potential to do real harm.
The potential for harm is minimally discussed in the works the authors are aware of, so we provide a discussion here.
We argue that responsible coordination detection can be done through analysis of synchronized actions, a tactic which is highly misleading to the general population and one unlikely to be used by genuine political organizers.
Based on this argument, we outline a Synchronized Action Framework which is general enough to encompass behaviors studied in previous work and provides a method for obtaining all the edges efficiently.
This allows for the study of multiple behaviors at the same time, and requires the actions be synchronized to minimize the harm of detecting users coordinating in an authentic manner. 
We demonstrate two general coordination detectors following the framework, and find complex synchronization behavior in the amplification of the Reopen America Protests.

\subsection{Contributions}
This work has two main contributions: a discussion of responsible coordination detection on social media, which previous studies had not explored; followed by our proposed Synchronized Action Framework for responsible detection of coordination.
Specifically, we study hashtag usage, URL usage and mention usage and their combinations, but this framework encompasses methods to study coordination along text, images, account creation times, and more.
Lastly, we demonstrate our method on a large twitter dataset.
The method is validated though the detection of 3 templated coordinated campaigns.
We find that covert coordination detection is far more complex than these simple examples or that seen in some prior works.
Still, we find a cluster of suspicious users, and detail 3 of their actions in amplifying the Reopen Protest rhetoric.

\section{Prior Work}
The field of social cybersecurity is exploding in interest as social media's role in our threatening peace reach the forefront of public discussion \cite{carley2018social,uyheng2019interoperable}. 
This is especially true following the recent Washington D.C. protest and subsequent breach of the nation's Capitol, which was fueled by mis/disinformation both on and off social media. While ``fake news" on social media has led to many real world consequences, there are still many unanswered questions \cite{grinberg2019fake}.
Here, we focus on one of these questions: How can we responsibly discover coordinated accounts?

One of the closest lines of inquiry is bot-detection, which seeks to identify automated users on social media \cite{ferrara2016rise}. Common bot-detection mechanisms are (1) feature based, using account details such as tweet frequency and content quality \cite{beskow2018bot,davis2016botornot}; or (2) graph-based, using an account's communication links for bot classification \cite{cao2012aiding,magelinski2020graph}. Feature-based scale better to large datasets, however they only consider accounts in isolation. 

The bot-detection algorithms that are closest to coordinated-account detection are DeBot and CopyCatch. DeBot \cite{chavoshi2016debot} finds correlated accounts on Twitter based on the temporal activity of user tweets, but does not account for tweet content, an essential component for detecting coordinated behavior. CopyCatch \cite{beutel2013copycatch} was built to detect coordinated attempts to inflate the ``likes" on Facebook Pages. While it can be generalized to detect synchronized actions, it is designed to detect coordination within one burst of time, rather than the long-term coordination we can detect via a network-based approach.

Recently, the gap between bot detection and coordinated activity has been recognized. \citet{pacheco2020uncovering} introduced a framework for analyzing coordination. 
They first define an action type that is coordinated, e.g. hashtag usage, and create a user-behavior bipartite network.
Finally, they construct a coordination network by folding the user-behavior network into a user-user network, with network links representing the strength of the shared behavior.
Typically, their networks are sparse enough to perform component analysis of coordinated users.
When component analysis is too coarse, \citet{nizzoli2020coordinated} gives a more principled procedure through network community detection algorithms.
\citet{vargas2020detection} furthered the use of network statistics in the prediction of coordinated activity on Twitter, defining coordination behaviors in terms of retweets, co-tweets, hashtags and so forth. 
Taking a different approach, \citet{sharma2020identifying} presents the problem as a temporal point process to model an account activity in terms of events.

These coordination frameworks only consider actions individually, despite a large body of work on multi-view network analysis \cite{aleta2019multilayer,cruickshank2020multi}.
Further, synchronized action is not built-in to the existing frameworks, and attempting to construct synchronized-action networks with these approaches results in losing up to half of the network's edges.
Most crucially, none of the works that we are aware of have provided substantial discussion of the ethical implications of the problem itself.
Here, we attempt to resolve the methodological issues, and provide a first discussion on responsible coordination detection.

\section{Responsible Coordination Detection}
\label{sec:ethics}
The detection of coordinated actors on social media is necessitated by the increasing interest in social media manipulation. However, to the best of our knowledge, none of the studies of coordinated behavior explore the ethical implications of this problem further than the acknowledgement that it is difficult to distinguish between ``good" and ``bad" coordination. We push this line of inquiry further, and borrow Kalluri's framing to ask how successful coordination detection may shift power \cite{kalluri2020don}. 
Note that we consider not yet realized ``successful" coordination detection, to limit sightedness \cite{russell2019human}.
Specifically, we do so in the context of protests, given the recent interest in coordinated activity surrounding the Reopen America protests and the Capitol Riot.

In the ideal case, coordination detectors are able to identify meaningful sets of users who are cooperating in some way on social media. Social media is an effective tool for alternative protest reporting, giving the public the power to directly report events as they experience them \cite{hermida2018twitter}. It provides the public a way to find and organize support for a cause, sometimes escalating into offline protests. 

At the same time, the surveillance of social media by law enforcement and government agencies across the word is widespread and still increasing \cite{mateescu2015social,patton2017stop,williams2013policing,qin2017does}, with efforts expanding from policing online actions to policing of protests and mobilization \cite{dencik2018prediction}. 
Social media information about protests and political dissent is not only used for general intelligence but has also been used for proactive disruption of organized protest through preemptive targeted arrests on key actors in the organizer networks \cite{dencik2018prediction,swain2013disruption}. 
Such use of social media against the public in protests have led many to conclude that the power of social media has shifted away from the public towards law enforcement and government agencies \cite{mateescu2015social}. 
In addition, this shift can be particularly negatively impactful to minority communities already facing a large power imbalance \cite{owen2017monitoring,qin2017does,patton2017stop}. 
Previous demonstrations and riots like the London and US Capitol riots have been used to justify further development in the technologies driving this power shift \cite{williams2013policing}, increasing pressure to push this imbalance even further.

Given that there is a large and potentially increasing desire for law enforcement and government agencies to understand social media, and that their existing power in this space has been used to disrupt political demonstrations, it is clear that unrestricted coordination detection increases existing imbalances in power against the public. 

Next, we narrow the view of the coordination detection problem to that of \textit{automated} coordination. 
The ethical implications of automated coordination are far different from general coordination.
Grassroots organizers are unlikely to be automating their messages or calls for protests, instead relying on a network of community members for a more personal outreach. 
An automated approach relies on inorganic amplification of ideas, messages or stories. 
This amplification can mislead ordinary social media users, who expect popular tweets to be popular with people, not automated accounts.
Detection of this automated behavior, then, shifts power away from those in control of networks of automated accounts and towards ordinary users, who can now have more trust in the content on their feeds.

In practice, coordination detection of all kinds must be held to similar standards as bot detection. There must be clear rules which are consistently and transparently enforced should an account be suspected to take part in coordinated activity, and platforms must give users the ability to appeal. This is especially important in coordination, since it is possible to make an account seem coordinated by building a network around it. Further, a coordination detection strategy should not be fully automated and should have human feedback loops to rectify mistakes.

We propose a Synchronized Action Framework that allows us to specifically target automated coordination. Synchronized action is very difficult to achieve with real human users, which has led to its successful application in high-precision bot detection and coordination detection \cite{chavoshi2016debot,pacheco2020uncovering}. Our framework is a general one that leverages on synchronized actions to detect many coordination strategies.

\section{Synchronized Action Framework}
\label{sec:methods}
A synchronized action occurs when two users take the same action at the same time.
We define an ``action'' as any measurable behavior a user exhibits that can be  localized to a point in time.
An example action is a user tweeting ``\#reopen.''
More generally, ``action-types" are sets of potential actions.
For example, hashtag usage is an action-type, while tweeting \#reopen, \#liberate, or \#openup are all different actions of the hashtag-type.

Action-types are used to define the views of the multi-view coordination network.
That is, per action-type, users are connected to other users based on the strength of their synchronization among actions within their type.
If hashtag usage is an action-type of interest, there will be a hashtag-view in the coordination network, connecting users based on how strongly their hashtag usage was synchronized.

The definition of synchronized action can be relaxed by considering \textit{similar} actions or \textit{similar} times.
DeBot, for example, looks to find users that have correlated tweet times \cite{chavoshi2016debot}.
Posing their approach under our framework, the action \textit{and} action type is simply to tweet, giving a one-view coordination network.
The measure of synchronization used is temporal correlation.
Thus the choice of action-types, synchronization measure and network analysis approach fully defines the method.

In this study, we use a relaxed \textit{same-action, similar-time} definition of synchronized action. We first consider 3 standard action-types: hashtag, URL and mention usages, resulting in a 3-view network.
We measure synchronization through a 5-minute sliding window described in detail in Section \ref{sec:view_construction}. Selecting a small window minimizes the connections between normal users who are taking actions at the same time as others by chance.
We then use multi-view clustering to find the cluster of highest density, and investigate central nodes within.
Lastly, higher-order action types of order two are considered.
Specifically, our 3 higher-order actions are: hashtag and URL, URL and mention, and hashtag and mention, all of which are only counted if both actions occur in the same tweet.
While these higher-order actions result in subsets of the original actions, the direct connection between content (hashtag or URL) and target (mention), makes discovered coordinated campaigns easy to analyze under the BEND framework \cite{carley2020social}.

\subsection{Multi-View Coordination Network}
A Multi-View Coordination Network is a $L$-layer network $G = \{V,E,\mathcal{L}\}$ where $\mathcal{L}$ is the set of view indices $\{1, 2.., L\}$. $V =\{V^1 \cup V^2 ... \cup V^L\}$, where $V^i$ denotes the set of nodes in layer $i$ of the network. 
$E = \{E^1 \cup E^2 ... \cup E^L\}$ where $E^i$ denotes the set of edges in layer $i$ of the network.
Specifically, nodes $V$ are Twitter users, edges $E$ are connections between users based on coordination actions. Each view $L^i \in \mathcal{L}$ represents an action-type $i$. 
Standard action-types are single instances like a common hashtag between two tweets. 
Higher order action-types are a combination of standard action-types: edges for the (hashtag-URL) action-type are formed when two tweets contain a common hashtag \textit{and} URL.

In each view $L^i$, the action-type $i$ can have multiple actions, $A^i = \{A^i_1,A^i_2,...A^i_k\}$. For example, the hashtag action-type encompasses many actions: \#reopenNY, \#reopenTX or \#reopenPA.
An edge $e_{x,y} \in E^i$ represents the presence of coordination action-type $i$ between users $x$ and $y$ within a time window $t$. 
The network can contain directed or undirected edges, or both. Weighted edges $e_{x,y}$ have a weight value $w_{x,y}>0$, which represents the strength of coordination between the two users.
Here, we consider undirected weighted edges.

In our first example, a multi-view coordination network is formed with three layers, or three standard action-types: $L=\{\text{hashtag, URL, mention}\}$. 
For example, if users $x$ and $y$ post tweets with the hashtag ``\#reopenNY'' within $t$, an edge  $e_{x,y} \in E^1$ is drawn.
In our second example, we form a multi-view coordination network with three higher-order action types: $L=\{\text{(hashtag-URL), (URL-mention), (hashtag-mention)}\}$. 
For this work, we set $t=5$ minutes.

\subsection{View Construction}
\label{sec:view_construction}
The bipartite-approach to synchronized action used by \citet{vargas2020detection} defines an action-type by taking the intersection of a previous action and a time period.
As such, the action of tweeting a hashtag becomes the action of tweeting a hashtag within a time-window.
Thus, a user-action bipartite network is created and folded to get a synchronized action network.
This approach fails to capture 50\% of a network's edges under a uniform distribution of actions in time. 
In practice, the proportion of missing links is less than 50\% but still large.

A naive approach to obtain all links is to sort tweets chronologically, move a sliding window across the tweets, then count instances of shared-behavior. 
Unfortunately, this approach presents an $O(N^2)$ time complexity, where $N$ is the number of tweets that require pairwise comparison. 
However, only tweets with the same action need to be compared.
Thus, we can group by action, and run the procedure on the subsets.
While the time complexity of the procedure is still $O(N^2)$, $N$ has now been reduced to the subset of tweets containing a specific action rather than the set of all tweets, improving the runtime.
Since the maximum number of tweets containing a specific action is typically far smaller than the set of all tweets, the sliding window approach goes from intractable to scalable.
This also implies that the more specific action types are more scalable.
For example, if the action type is the most general action possible, simply tweeting, $N$ is still the set of all tweets.
A very specific action type like tweeting both a hashtag and a URL will greatly reduce $N$.

Further, decreasing the time window increases scalability. 
The quadratic comparison operation only takes place for tweets falling within the sliding time window of interest.
In the worst case, all of the tweets fall within the time window, resulting in $O(N^2)$.
However decreasing the size of this window decreases the number of necessary comparisons.
Thus, the small time-scales of inauthentic coordination are ``built--in" to our method, and not something that simply be changed with a parameter to find more authentic coordination.

When a sliding window is used, as opposed to time-segments, the weighting of connections becomes non-trivial.
This is particularly important for coordination, where erroneously high edge weights can result in normal users looking suspicious.
In a static window, the edge weight between two users should correspond to the minimum number of instances between two users in the window.
For example if user A is in the window twice, while user B is in the window 100 times, the connection should only be of strength 2.
Otherwise, spammers will be erroneously strongly connected to normal users.
In a sliding window, this is more complex.
We are considering a single user which is about to be moved out of the window, and we have to decide which users to draw connections to.
Unless all prior connections are recorded, there is no way of knowing if a connection between two users was already in one of the last $m$ window moves.
To account for this, we only draw connections from the user in question to others if the user has a lower presence in the time window than the other potential connecting user.
This procedure behaves similarly to the analogous fixed-window approach.

\subsection{Multi-View Network analysis}
We first cluster the multi-view network with the multi-view modularity clustering technique \cite{cruickshank2020multi}. Multi-view clustering seeks to combine different divisions, or views, of the data to produce a better clustering output. 
There are a number of ways to analyze the resulting clusters.
Here, we study the cluster with the highest density.

We then explore nodes with highest centrality in terms of total degree, eigenvector centrality, and community-hub modularity vitality \cite{magelinski2021measuring}.
Total degree is of particular importance, as it gives the nodes with the highest number of coordinated connections between users which are the most likely to truly be coordinating.

\subsection{Action-Type Selection}
We begin by selecting standard action-types on twitter: sharing hashtags, sharing URLs, and mentioning other users.
However, these action-types can lead to false-positive connections. This is particularly true in the case of hashtags, where some users were observed to be strongly connected but actually tweeting opposing views.
As such, singular action-types of hashtag and URL usage may be too general.

To remedy this, we combine the standard action-types into  higher-order actions of order two.
For example for the (hashtag-URL) action-type, we consider tweets which contain both a hashtag \textit{and} a URL. 
Tweets that contain multiple hashtags and URLs are separated into all possible (hashtag, URL) tuples.
The notion of higher-order action leads to more specific behavior, with less likelihood of making erroneous user connections.
This also leads to a scalability advantage.
The definition of very specific behavior within a short sliding time window leads to few but meaningful edges in the coordination network graph, making the network analysis easier, as has been shown in the prior work on coordination detection. 

\section{Results}
In this section, we present the results of coordination detection on the Reopen America Twitter dataset.
We first demonstrate the approach's validity though detection of three coordinated campaigns.
Lastly, the added complexity of \textit{covert} coordination operations is illustrated, and our framework's effectiveness in finding coordinated users under this complexity is demonstrated.

The Reopen America protests occurred across the United States from April to September 2020,  with protesters hoping to lift the COVID-19 safety restrictions affecting daily work and recreation. 
These protests were widely regarded as dangerous to public health due to the disregard for social distancing and mask protocols. 
The similarity of protest organizing materials, dates, and phrasing across states raised questions about coordination, making this an interesting dataset to examine.

\subsection{Data}
By leveraging Twitter Search API, a large corpus of tweets from April $1^\text{st}$ 2020 to June $22^\text{nd}$ 2020 was collected.
The tweets were collected by searching the keywords and hashtags with ``openup", ``reopen", ``operationgridlock", and ``liberate". 
Additionally, all of the US State abbreviations were appended to each of the search terms, e.g. ``liberateNY", to collect the Reopen America protest for the corresponding state. 
As we will demonstrate, these search terms capture tweets about the protests, but also tweets about other topics which we find coordinated activity in.
In total, the dataset contains roughly 3.6 million unique users and 9.9 million tweets.

\subsection{Validation}
\label{sec:validation}
The effectiveness of our framework is now demonstrated though a number of examples. 
First, the ``standard-action" analysis is considered, where we discover the 7News network. 
We then move to higher-order action analysis to find more specific campaigns.
Unlike the previous analysis, we see little connection between views indicating that very few coordinated clusters use all three actions.
The multi-view cluster of note is the same 7News network.
The other strong connection across views belong to two affiliated accounts promoting a Facebook group to discuss school reopenings.
Given the disconnect between views, we proceed with individual view analysis.
In this analysis, we discover two additional campaigns: the ``Mexico Without Plastics" campaign, and the campaign to reopen the office-transfer service for tax inspectors in India.

\subsubsection{7News Network}
Before clustering, we observe many small components of the multi-view network, each containing publicly-affiliated accounts such as news networks.
We also examining the strongest link in the network, averaged across views, and find these two accounts are also formally affiliated.
This preliminary analysis provides strong face-validity.

\begin{figure}[!htb]
    \centering
    \includegraphics[width=\columnwidth]{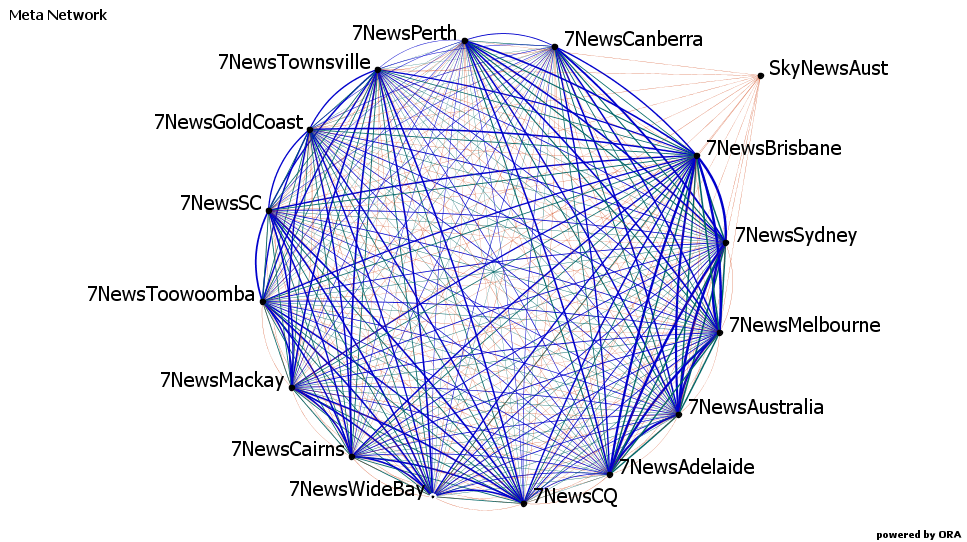}
    \caption{7News Component of the Multi-View Coordination Network. Link width corresponds to coordination strength. Blue, red, and green correspond to URL, mention, and hashtag coordination, respectively.}
    \label{fig:7news}
\end{figure}

The most well-connected of these components is the Australian 7News network, shown in Figure \ref{fig:7news}.
These accounts consistently rank highest in degree centrality and in community-hub modularity vitality.
The network is highly coordinated along all action-types: $A^{\text{hashtag}}$, $A^{\text{URL}}$ and $A^{\text{mention}}$.
The accounts tweet the same stories at the same time, often mentioning the reporter, using the URL of the story, and adding hashtags like \#7News.
They appear in our dataset for their coverage of ``reopen" events, mostly focusing on the eventual reopening of Australian borders.
The coordination between local accounts broadens their geographic coverage, while also reinforcing the popularity of their URLs and hashtags.

\subsubsection{Mexico Without Plastics Campaign}
\begin{figure}[!htb]
    \centering
    \includegraphics[width=0.8\columnwidth]{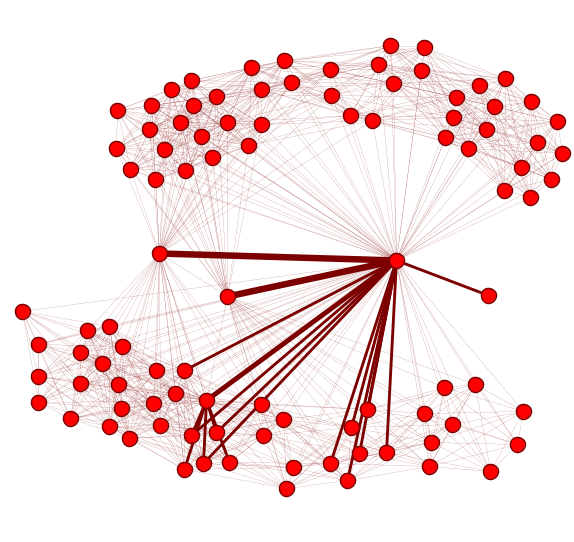}
    \caption{Ego of the URL-Mention coordination network of the most-connected user in the Mexico Without Plastics Campaign. Link edges are weighted on a log-scale. Account names are redacted.}
    \label{fig:mexico}
\end{figure}

Investigating the strongest connections in the $L^{\text{URL-mention}}$ view of the higher-order network uncovers the ``Mexico Without Plastics" campaign to ban single-use plastics.
This campaign was sponsored at least in part by Greenpeace Mexico, hosting a petition calling for Mexico's senate to reform the waste-management laws in order to ban single use plastic.
The petition garnered over 275,000 signatures.
The ego network of the most-active account is shown in Figure \ref{fig:mexico}.
We see that a few coordinating actors were key leaders in promoting the campaign.
In this campaign, the URL is the petition, while the mention is either the official account of the Mexican senate, or its members. The campaign is present in our dataset due to the keyword ``lib\'erate," as in free yourself from plastic.

\subsubsection{Campaign to Open ICT for Indian Tax Inspectors}
Lastly, the $L^{\text{hashtag-mention}}$ view of the higher-order network is studied.
Here there is a dense cluster of users lacking the ``hubs" seen in the Mexico Without Plastics campaign, suggesting a lack of centralized leadership.
These users are campaigning in India for inspectors in the tax office (CBIC) to be granted the ability to have internal transfers (ICT).
Thus, the users are coordinating with hashtags like \#DepressionKillsOpenICT and \#CGSTinspectorsWantICT, while mentioning Prime Minister Modi, government officials, and news outlets.
The campaign appears in our dataset form the phrase ``\textit{reopen} ICT."

\subsection{Coordinated Amplification of Reopen Protests}
While the previous examples demonstrate validity of our approach in detecting coordinated campaigns, the campaigns themselves are templated examples.
They are intentionally public-facing, clearly coordinated and result in strong network structures like disconnected components and extremely strong edges.
The coordinated activity surrounding the Reopen America protests is less clear-cut and more interesting, showing the difficulty of detecting covert coordination.

We find strong connections within the densest cluster of the first-order multi-view coordination network, as found through multi-view modularity clustering.
We study some of the strongest connections in this cluster.
The most central node in terms of total degree centrality is FedUpUSA, a self-proclaimed media account committed to bringing ``truth about what is really happening, as opposed to the fodder that is shown in the mainstream media". 
Their website has been inactive for 2 years, while their twitter account remains active.

\begin{figure}[!htb]
    \centering
    \includegraphics[width=\columnwidth]{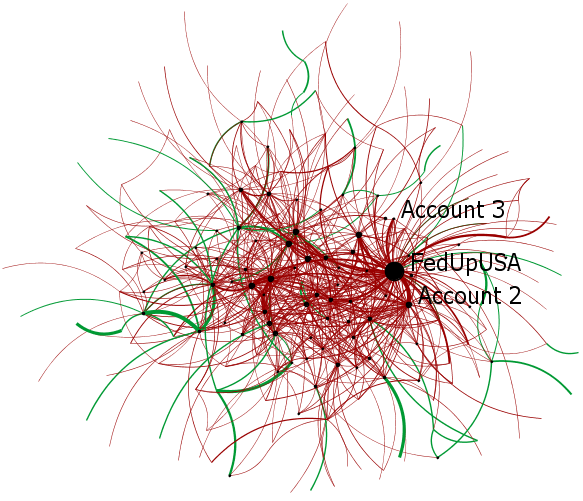}
    \caption{FedUpUSA multi-view coordination ego network. Red, and green correspond to mention, and hashtag coordination, respectively. Link width represent normalized mention connections. Nodes are sized by their total degree.}
    \label{fig:fedup}
\end{figure}

The multi-view ego network of FedUpUSA in Figure \ref{fig:fedup} demonstrates its complexity. 
In comparison to previous campaigns, we see the coordinated activity around the reopen protests to be far murkier.
The lack of simple coordination is seen though the absence  of any obvious strongly connected clique, indicative of either a lack of coordination or a more sophisticated strategy.
Thus, network measures and account content analysis becomes even more important than they were in the simple examples seen in the validation section and in prior works.

Since FedUpUSA is strongly connected to a number of accounts, we focus on the top-2 alters here, though anecdotally many of the strongest alters display similar behavior.
Both users appear to be private citizens or have protected accounts, so we redact the handles here and refer to them as Accounts 2 and 3 going forward.

All accounts were highly active during the protests, but each had a different regional focus.
FedUpUSA focused on Michigan, Account 2 on Minnesota, and Account 3 on North Carolina. 
All accounts pushed hashtags related to the protests, with the most popular being  \#OperationGridlock, \#EndTheShutdown, each being used over 40 times combined between the users, not counting retweets. 
FedUpUSA was particularly focused on hashtag usage, often tweeting pictures of the Michigan protests with up to 10 hashtags while mentioning news outlets and public officials.
Account 2 tweeted and retweeted supportive statements about the Michigan protests, before focusing on Minnesota.
The Minnesota tweets included times and locations of one protest, live-stream information, and supportive statements from public figures such as the President.
Account 3 did not focus on the protests themselves, but instead on sending the ``reopen" message directly to the North Carolina Governor and other officials through mentions.
They frequently used hashtags such as \#ReopenNC, and \#VoteCooperOut.
They also focused on reopen ``successes," such as a court order allowing religious organizations to meet.

All three accounts produced unique content and interacted genuinely with other accounts, indicating that bot-detection methods would not identify them, highlighting the distinction between coordination-detection and bot-detection.

While these accounts had strong overlap in hashtag and mention usage, they did not share any of the same URLs, highlighting the difference in coordination strategies between groups such as the 7News organization, which relied heavily on URLs.

\section{Limitations}
First and foremost, our notion of coordination is highly limited.
We intentionally limit the type of coordination we investigate to synchronized action due to the ethical considerations discussed in Section \ref{sec:ethics}.
However, there are many other methods of coordinating which we are not looking for.
More importantly, without knowledge of private affiliations or plans, we cannot truly say that accounts are coordinating.
Instead, our methods point to suspicious users, which are taking the same actions at the same time.
It is entirely possible that the users we found to be synchronized during the reopen protests did not in fact work together behind the scenes, so we make no claims about direct account affiliation when they are not made clear by the users themselves.
However, we note that regardless of affiliation, their impact is similar since they pushed similar messages at similar times.

As can be seen in Figure \ref{fig:mexico}, synchronization-based networks create network chains, wherein many users are weakly connected because their tweets happen to occur at a similar time to many others.
This raises a challenge in differentiating coordinated efforts from organically emerging support.
In this work, we tested a weighting scheme, wherein popular actions are down-weighted.
This weighting scheme was successful at minimizing ``false positive" connections, though it obscured overall network structure.
Thus, weighting schemes which both minimize inadvertent connections while retaining coordination structure is of high importance for future work. 

While the coordinated actors we discovered in this work do appear to be spreading the same messaging, this is not guaranteed, especially under single-view analysis. 
Hashtags in particular can serve multiple roles, so a hashtag-only analysis could be connecting users with opposing views on a hashtag that are talking at the same time \cite{yang2012we,magelinski2020canadian}.
Further, we may be inadvertently detecting ``hashtag hijacking", i.e. using popular hashtags to increase viewership of their tweets. 
We believe this is mitigated through the other action type connections and the focus on strong links, though future work leveraging tweet text will be even stronger.

Our dataset is a primarily English dataset from Twitter, due to the chosen keywords. 
While we see coordination among English speakers (reopen protests) and Spanish speakers (plastics campaign), our results are still highly skewed towards users tweeting in English.
Future work could explore collection of the same hashtag in multiple languages to build a coordination network across different languages.
Additionally, keyword-based collection often retrieves tweets that are not related to the event itself, hence resulting in our detection of unrelated campaigns, like the Indian Tax campaign.

The collection strategy also creates a bias in the tweets for $L^{\text{hashtag}}$ and against $L^{\text{URL}}$.
By collecting tweets with keywords, we are restricted to finding URLs which either contain the terms themselves, or are posted along with one of the keywords.
The significantly higher number of connections in the hashtag view compared to the link view may effect the quality of the final clusters, since the current techniques in multi-view clustering work best under similar view-statistics.


\section{Conclusion}
Coordination detection has the power to do real harm, however we believe that responsible detection of coordinated activities on social media can be done through analysis of synchronized actions. We therefore propose a multi-view Synchronized Action Framework, enabling the analysis of coordination behavior across several actions simultaneously.
This general framework encompasses behaviors studied in previous works, and allows for expansion into others.

Our approach's effectiveness is shown using the Reopen America dataset.
Three templated coordinated activities are found: The 7News Network's effort to further their content though locally-based accounts tweeting together, the ``Mexico Without Plastics" campaign to reform single use plastics through waste legislation, and the ``Reopen ICT" campaign to reopen transfers for Indian Tax Inspectors.
These examples are publicly coordinated and simple to analyze, mirroring those seen in prior work, where network analysis is straightforward.

This contrasts starkly with the covert coordination in the Reopen America protests, which is complex and difficult to analyze.
With this added complexity, the multi-view approach we introduced becomes necessary.
We find a cluster of suspicious accounts and detail the actions of users who pushed similar hashtags and mentioned the same users at similar times while emphasizing protests in other states individually.

The complexity of covert coordination seen in this work highlights the need for a multi-action approach, where even more actions are considered.
New actions may be tweet text, images, account creation, and more.
Further, analysis of different time-scales may be insightful.
Our Synchronized Action Framework encompasses all of these avenues of future work.

\subsubsection*{Acknowledgments}
We would like to thank Matthew Babcock, PhD for collecting the Reopen America Twitter dataset used for this study. This work was supported in part by the Office of Naval Research (ONR) Award N000141512797 Minerva award for Dynamic Statistical Network Informatics, and the Center for Computational Analysis of Social and Organization Systems (CASOS). Thomas Magelinski was also supported by the IDeaS Center as a Knight Fellow. The views and conclusions contained in this document are those of the authors and should not be interpreted as representing the official policies, either expressed or implied, of the ONR.

\bibliographystyle{named}
\bibliography{references}{}

\end{document}